\documentclass[conference]{IEEEtran}
\IEEEoverridecommandlockouts

\usepackage{cite}
\usepackage{url}
\usepackage{amsmath,amssymb,amsfonts}
\usepackage{algorithmic}
\usepackage{graphicx}
\usepackage{textcomp}
\usepackage{mathrsfs}
\usepackage{multirow}
\usepackage{xcolor}
\usepackage{booktabs}
\usepackage{stfloats}
\DeclareUnicodeCharacter{FB01}{fi}

\newcommand{\tabincell}[2]{
\begin{tabular}
{@{}#1@{}}#2
\end{tabular}}

\begin{document}
%
\title{Blockchain Based Zero-Knowledge Proof of Location in IoT}



%
\author{\thanks{
    This work is supported in part by the grants from the National Science Foundation of China (No. 61571330), Shanghai Integrated Military and Cilivian Development Fund (No. JMRH-2018-1075), and Science and Technology Commission of Shanghai Municipality (No. 19511102002). Corresponding author: Erwu Liu.
}\IEEEauthorblockN{Wei Wu, Erwu Liu, Xinglin Gong and Rui Wang}

\IEEEauthorblockA{College of Electronics and Information Engineering, Tongji University, Shanghai, China \\
Emails: \{1832912, erwuliu, xinglingong, ruiwang\}@tongji.edu.cn}}


\maketitle

\begin{abstract}
With the development of precise positioning technology, a growing number of location-based services (LBSs) facilitate people's life. Most LBSs require proof of location (PoL) to prove that the user satisfies the service requirement, which exposes the user's privacy. In this paper, we propose a zero-knowledge proof of location (zk-PoL) protocol to  better protect the user's privacy. With the zk-PoL protocol, the user can choose necessary information to expose to the server, so that hierarchical privacy protection can be achieved. The evaluation shows that the zk-PoL has excellent security to resist main attacks, moreover the computational efficiency is independent of input parameters and the zk-PoL is appropriate to delay-tolerant LBSs.
\end{abstract}


\section{Introduction}
In recent years, the popularity of smart terminal devices and the development of sophisticated high-precision position sensors have led to many location based-services (LBSs) \cite{junglas2008location}, such as location-based rewards and digging services, location-based recommendations and location-based social network (LBSN) \cite{jiang2014locations}. Most LBSs require the user's proof of location (PoL) to prove that the user satisfies service requirements. For example, a user needs to prove that he is in the mart to obtain a discount coupon. While enjoying the convenience of these services, the user's location privacy is always revealed through the PoL to the server or malicious adversaries. Hence, when the user provides PoL to the server, the user's privacy must be protected.

A large number of related works on PoL emerge to protect the user's privacy. The previously proposed PoL system architectures can be roughly divided into two types: centralized and distributed. In the former, the user data is stored in the central server, while the latter one applies distributed storage. Tyagi et al. \cite{tyagi2015location} propose a location privacy architecture based on the approach like a mix zone model with other approaches to protect location privacy. Javali et al. \cite{javali2016alice}, based on channel state information (CSI) and fuzzy vault, a cryptographic primitive, propose a location proof generation and verification scheme in which provers can prove their presence in particular time with high confidence. Li et al. \cite{li2015privacy}, relying on the existing WiFi or cellular network access points (APs), propose an infrastructure-based scheme to provide privacy-preserving location proof, in which the database can verify the location without knowing the user’s accurate location. All the above solutions perform in a centralized way, which would lead to the following shortcomings. First, a powerful dedicated central server is needed to serve a large number of nodes, which results in an unbalanced load between the server and the node. Once the server is overloaded, the entire system will be paralyzed. Second, a malicious adversary is easier to capture the system because it has a single opponent, the central server. Once the central server is captured, the entire system is captured. Third, the user's data is stored in the central server and only the server can control them. This will lead to the user's losing control of his data.

In order to solve the problem of centralized architecture, Amoretti et al. \cite{amoretti2018blockchain} propose a blockchain-based PoL system, in which users access a blockchain network in a self-organized mode so that the users can communicate with neighboring $witnesses$ and request them to issue location certificates.

Although this solution does not require a dedicated IoT device, the security will be compromised because it does not restrict mobile nodes' access to the network and the permission of the nodes already accessed in the network is also unrestricted. Therefore, combined with blockchain technology and zero-knowledge proof, we propose a zero-knowledge proof of location (zk-PoL) protocol. Our zk-PoL protocol consists of two parts: the generation scheme and verification scheme. By generation scheme, an approved user could obtain a location certificate from the smart AP. Here 'smart' means that smart contact can be embedded in the AP to regulate AP's behavior and process data. By verification scheme, the user could use the location certificate to generate a zero-knowledge proof of location for LBSs he needs. Moreover, this zk-PoL protocol allows the user to achieve hierarchical privacy protection by generating different proofs with different input parameters. We also analyze the performance of the zk-PoL protocol in security and computational efficiency. The security analysis shows that the zk-PoL is able to resist the main attacks. The experiment demonstrates that the system performance is independent of input parameters and the zk-PoL is applicable to delay-tolerant LBSs. 

The remainder is organized as follows. In Section \uppercase\expandafter{\romannumeral2}, we introduce the essentials of blockchain and zero-knowledge proof. In Section \uppercase\expandafter{\romannumeral3}, we build our system model and based on it, we give our zk-PoL protocol. In Section \uppercase\expandafter{\romannumeral4} we analyze the security and computational efficiency of our system. In Section \uppercase\expandafter{\romannumeral5}, we conclude the work of this paper.
\section{Preliminary}
This section briefly introduces the knowledge of blockchain technology and zero-knowledge proof.
\subsection{Blockchain Technology}
\begin{figure}[ht]
\centering
 \includegraphics[width=6.5cm]{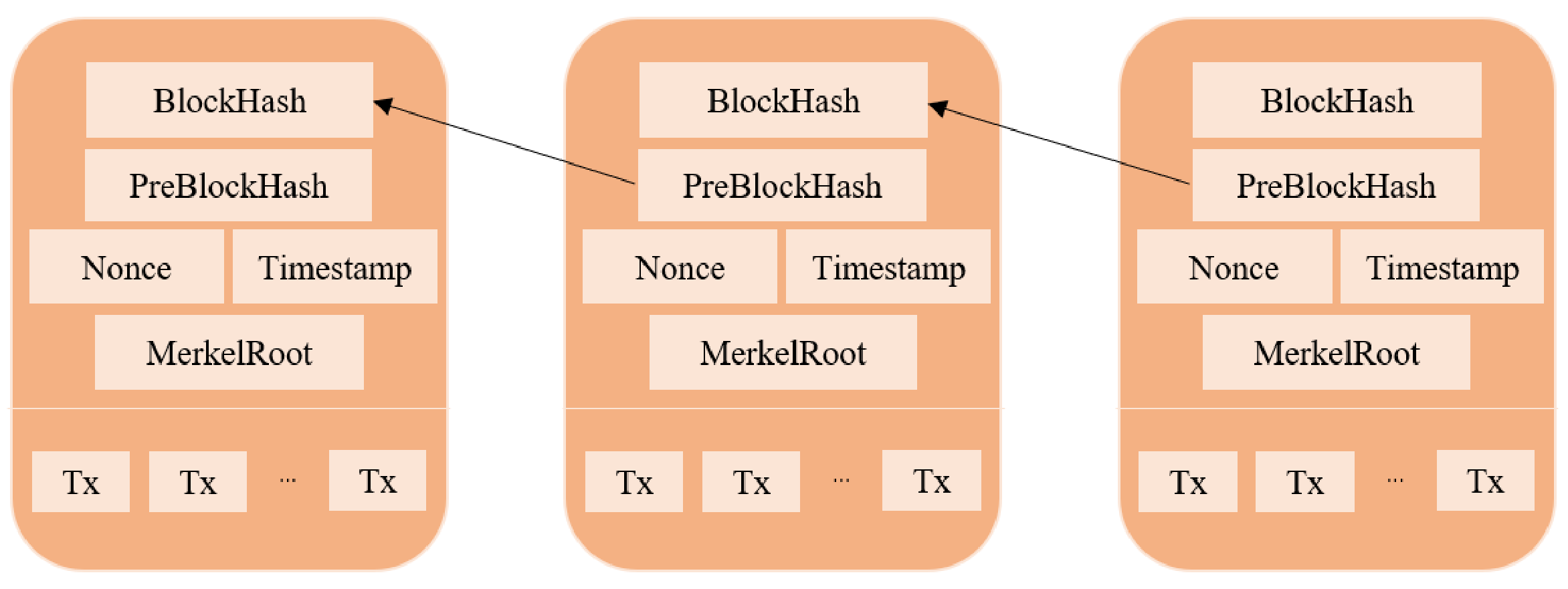}
\caption{Structure of A Simplified Block Chain}
\label{simplified block chain}
\end{figure}

Blockchain technology is a fusion of multiple technologies, including encryption, consensus mechanisms, incentives, distributed storage and smart contracts \cite{tschorsch2016bitcoin,nakamoto2008bitcoin,Ethereum}. In essence, a blockchain is a chained public ledger guaranteed to be indestructible with cryptographic methods. It consists of a series of blocks, and the latter block is connected to the previous one by including the cryptographic hash of the previous block. In addition, the block also contains timestamps, nonce, transaction data, etc. Its structure is shown in Fig.~\ref{simplified block chain}.
Then, by the consensus algorithm, such as Proof of Work (PoW) and Proof of Stake (PoS), all peers in the network agree on the ownership of the bookkeeping right. The peer that obtains the bookkeeping right adds the packaged block to the chain afterward. The data recorded in the chain cannot be modified. This whole process is defined as mining, and the peers participating in the bookkeeping right competition are defined as miners. For the public chain, the miner that finally obtains the bookkeeping right and completes the bookkeeping will receive a token reward. This mechanism is referred to as the incentive mechanism. Currently, there are at least four types of blockchain networks: public blockchain, private blockchain, consortium blockchain, and hybrid blockchain. The blockchain network introduced in this paper is the hybrid chain, in which different peers have different permission according to their diverse responsibilities.
\subsection{Zero-Knowledge Proof}
The zero-knowledge proof is a kind of method by which a $prover$ can convince a $veriﬁer$ that some statement is true without revealing any other information. Any zero-knowledge proof protocol should satisfy three properties \cite{Petkus2019WhyAH}:
\begin{itemize}
    \item Completeness: If the statement is true then a $prover$ can convince a $veriﬁer$.
    \item Soundness: A cheating $prover$ can not convince a $veriﬁer$ of a false statement.
    \item Zero-knowledge: The interaction only reveals whether a statement is true or not and nothing else.
\end{itemize}

Zero-Knowledge Succinct Non-Interactive Argument of Knowledge (zk-SNARK) is a novel form of zero-knowledge cryptography which is introduced in \cite{bitansky2012extractable} and built in \cite{groth2010short}. Quadratic Arithmetic Program (QAP) \cite{gennaro2013quadratic} and the Pinocchio protocol \cite{parno2013pinocchio} make it more effective and universal. Its first widespread application is Zcash \cite{sasson2014zerocash}. The following is a brief introduction to the zk-SNARK protocol \cite{Petkus2019WhyAH}.

Any complex computation or program can be regarded as one consisting of a series of simple operations which are defined as
 \vspace*{-0.5\baselineskip}
\begin{equation}\label{operatoin}
    left\ operand\ \textbf{operator}\ right\ operand=output,
 \vspace*{-0.5\baselineskip}
\end{equation}
where ${ operator }\in \left\{ +,-,\cdot ,\div  \right\} $. Naturally, if we would like to verify the correctness of the entire computation result, we just need to verify the correctness of the results of each operation separately, but such verification is very inefficient. Therefore, we can interpolate these operations into a polynomial to verify all operations at a time, that is, to verify polynomial
 \vspace*{-0.5\baselineskip}
\begin{equation}\label{computation polynomial 1}
    L(x)\ \textbf{operator}\ R(x)=O(x),
 \vspace*{-0.5\baselineskip}
\end{equation}
where $L(x)$, $R(x)$ and $O(x)$ represent left, right and output polynomials, respectively. This is the basic idea of QAP. 
Here, the \textbf{operator} is rewritten as a multiplication since the addition and subtraction are made in each operand, and the division can be converted into multiplication. It assumes that the computation can be decomposed into $d$ operations. If the computation is correct, the equation is true for each $x=i,\ i\in\left\{1,...,d\right\}$. Therefore, polynomial $L(x)\cdot R(x)-O(x)=0$ should have roots $x=i$, $i\in \left\{ 1,...,d \right\} $, which is denoted as 
 \vspace*{-0.5\baselineskip}
\begin{equation}\label{computation polynomial 2}
    p(x)=L(x)\cdot R(x)-O(x)=t(x)\cdot h(x),
 \vspace*{-0.5\baselineskip}
\end{equation}
where $\ t(x)=\prod_{i=1}^{d}{(x-i)}$.

In this way, by checking whether the polynomial $p\left(x\right)$ has the factor $t\left(x\right)$, the correctness of the computation can be checked. If a $prover$ intends to prove the correctness of the computation, he only needs to provide $L(x)$, $R(x)$, $O(x)$ and $h\left(x\right)$ which satisfy \eqref{computation polynomial 2}. In order to remain the same variable assignments of the same operands (or output) in different operations unchanged, $L(x)$, $R(x)$ and $O(x)$ are rewritten as the sum of the variable polynomials
\begin{equation}\label{assign}
    \left\{
    \begin{matrix} L(x)=\prod _{ i=0 }^{ n }{ { v_{ i } }{ l_{ i } }(x) }  \\ R\left( x \right) =\prod _{ i=0 }^{ n }{ { v_{ i } }{ r }_{ i }\left( x \right)  }  \\ O(x)=\prod _{ i=0 }^{ n }{ { v_{ i } }{ o }_{ i }(x) }  \end{matrix}
    \right.,
\end{equation}
where $i\in\left\{0,...,n\right\}$. $l_i(x)$ is the $i$th variable polynomial of the left operand; $n$ is the number of variables involved; $l_0(x)$ is variable polynomial of pseudo-variable $one$. The variable polynomials are determined by the specific computation. The $prover$ cannot modify these variable polynomials but only assign them, that is, multiply them with $v_i$. Knowledge-of-Exponent Assumption (KEA) is utilized to construct variable consistency polynomial. Variable consistency polynomial ensures variable assignments consistency across operands in the same operation, and it is denoted as
\vspace*{-0.5\baselineskip}
\begin{equation}
\begin{aligned}
   Z\left( x \right) =&{v}_{i}\sum _{ i=1 }^{ n }{ \beta _{ l }l_{ i }\left( x \right) +\beta _{ r }{ r_{ i } }\left( x \right) +\beta _{ o }{ o_{ i } }\left( x \right)  } \\ =&{ \beta _{ l } }L(x)+{ \beta _{ r } }R(x)+{ \beta _{ o } }O(x),
\end{aligned}
 \vspace*{-0.5\baselineskip}
\end{equation}
where $\beta_l$, $\beta_r$ and $\beta_o$ are random numbers.
Next, the $prover$ has to convince the $verifier$ that he knows the variable assignments that satisfy the restrictions, but the $prover$ cannot directly inform the $verifier$ of the variable assignments that satisfy \eqref{computation polynomial 2}. One reason is that it will expose all information to $verifier$, the other is that the multiplication of polynomials is extremely complex. According to the Fundamental Theorem of Algebra, for any polynomial, we can use its evaluation at any point to represent its unique identity. Therefore, the $verifier$ can verify whether the evaluation of \eqref{computation polynomial 2} is satisfied at a random point $s$, that is, verify
 \vspace*{-0.5\baselineskip}
\begin{equation}\label{encrypted 2}
    L\left(s\right)\cdot R\left(s\right)-O\left(s\right)=t\left(s\right)\cdot h\left(s\right).
 \vspace*{-0.5\baselineskip}
\end{equation}

\begin{figure}[ht]
\centering
\includegraphics[width=6.5cm]{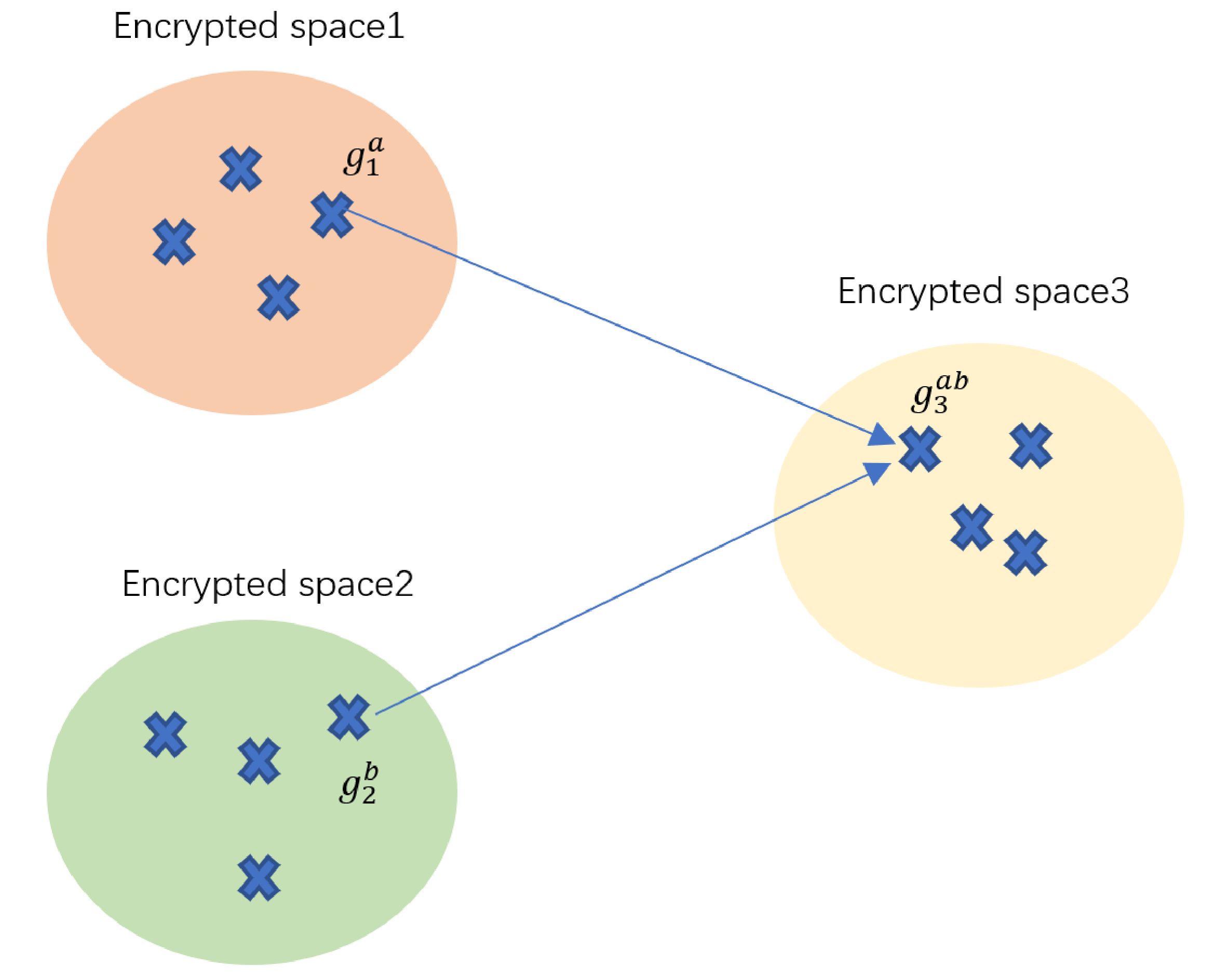}
\caption{Schematic Diagram of Bilinear Map}
\label{bilinear map}
\vspace*{-0.5\baselineskip}
\end{figure}

In order to ensure that the $prover$ cannot forge the polynomial, the random point $s$ should be encrypted. So, the $verifier$ needs to verify \eqref{encrypted 2} in the encryption, that is, to verify
 \vspace*{-0.5\baselineskip}
\begin{equation}\label{7}
    e(g^{L(s)},g^{R(s)})=e(g^{t(s)},g^{h(s)})\cdot e(g^{O(s)},g).
 \vspace*{-0.5\baselineskip}
\end{equation}
In addition, the $verifier$ also has to validate the constructing of polynomials by checking
\begin{equation}\label{8}
    \left\{
    \begin{matrix} \begin{matrix} e(g^{ L(s) },g^{ \alpha _{ l } })=e(g^{ { \alpha _{ l } }L(s) },g) \\ e(g^{ R(s) },g^{ \alpha _{ r } })=e(g^{ { \alpha _{ r } }R(s) },g) \end{matrix} \\ e(g^{ O(s) },g^{ \alpha _{ o } })=e(g^{ { \alpha _{ o } }O(s) },g) \end{matrix}
    \right.,
\end{equation}
\begin{equation}\label{9}
    e(g^{L(s)},g^{\beta_l})\cdot e(g^{R(s)},g^{\beta_r})\cdot e(g^{O(s)},g^{\beta_o})=e(g^{Z(s)},g),
\end{equation}
where $g$ is the generation point of elliptic curves and $e(g^*,g^*)$ is cryptography pairings (or bilinear map), which maps two encrypted inputs from two encrypted space to their multiplied form in different encrypted space, as shown in Fig.~\ref{bilinear map}.
\vspace*{-0.5\baselineskip}

If \eqref{7}, \eqref{8} and \eqref{9} are satisfied, the $verifier$ can believe that the $prover$ indeed knows true variable assignments with great confidence and do not know any other information about these variable assignments.

\section{System Model}
\begin{figure}[ht]
\centering
\includegraphics[width=8.5cm]{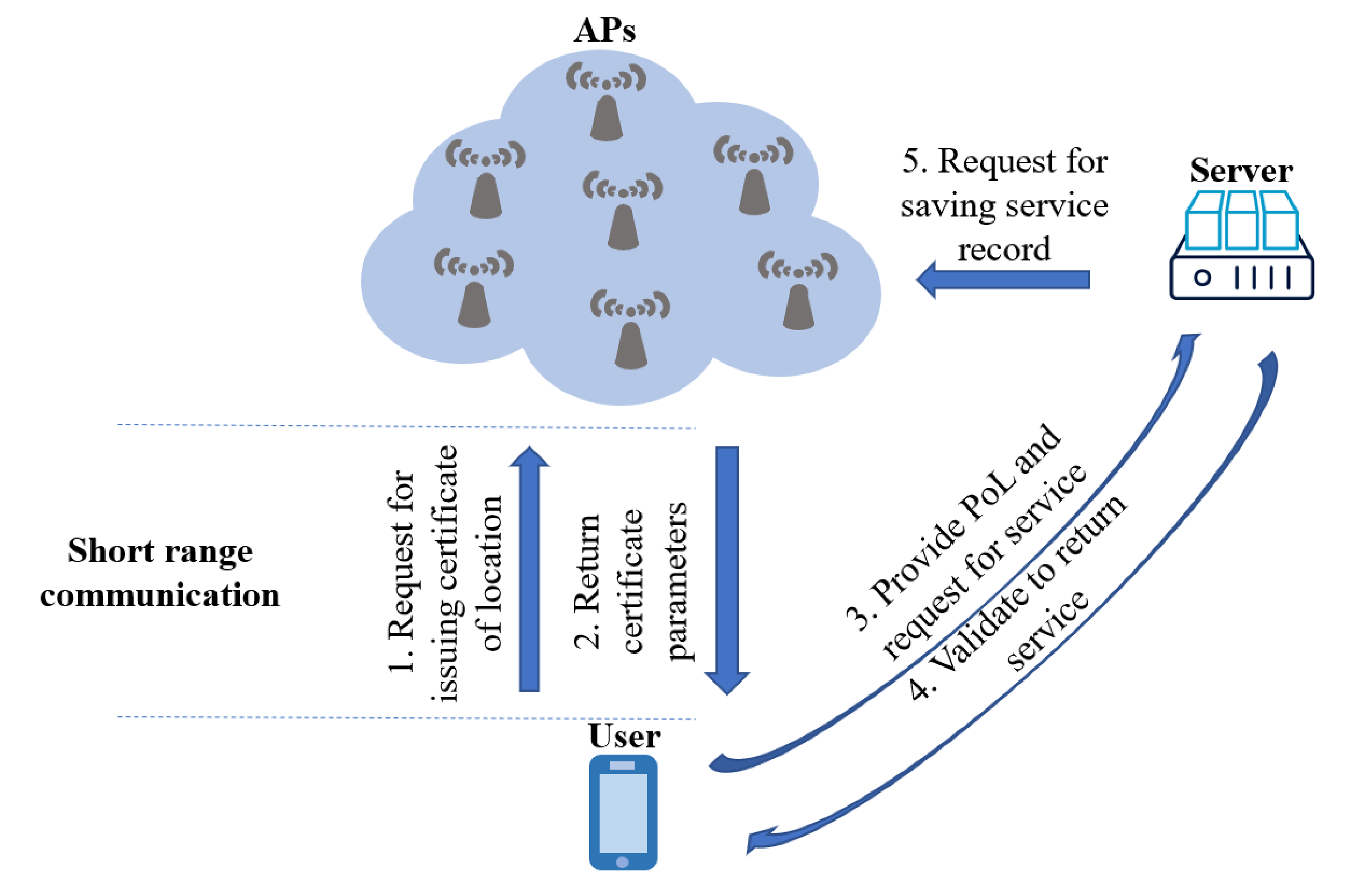}
\caption{System Model of zk-PoL}
\label{network architecture}
\end{figure}

The location data depict someone’s digital trace which gives more contextual information such as individuals’ habits, interests, activities and relationships. This information exposes the user to an unwanted advertisement and location-based spams/scams, causing social reputation or economic damage and making them victims of blackmail or even physical violence \cite{shokri2011quantifying}. Moreover, location data are insecurely transmitted and stored, and uncontrollable by its owner in traditional mode. In order to solve these drawbacks, we consider a system model with smart IoT devices. Our system model contains three components: the server, the user, and the ledger, as shown in Fig.~\ref{network architecture}, where the ledger is a blockchain network consisting of smart access points (APs). The general function of each component is described below.

APs are responsible for issuing location certificate to the user, providing the user with Common Reference String (CRS) for generating the proof and saving the service record, i.e. which service has been responded.

The user requests the nearby AP to issue a location certificate with short-range communication mode. When some LBS is needed, the user uses the location certificate to generate a proof for this LBS.

The server receives the service request and the proof sent by the user. After validating the proof, the server provides the user with the corresponding LBS. Once the service is obtained by the user, the server will request the AP to save the service record, updating the ledger.

This kind of system enables separation of the user's data and the service. The public ledger only holds the root of trust of data, and the actual data is stored locally by the user. When the user requests for LBSs from the server, he just needs to prove the truth and validity of location data that he holds.

\begin{figure*}[ht]
\centering
\includegraphics[width=16cm, height=7cm]{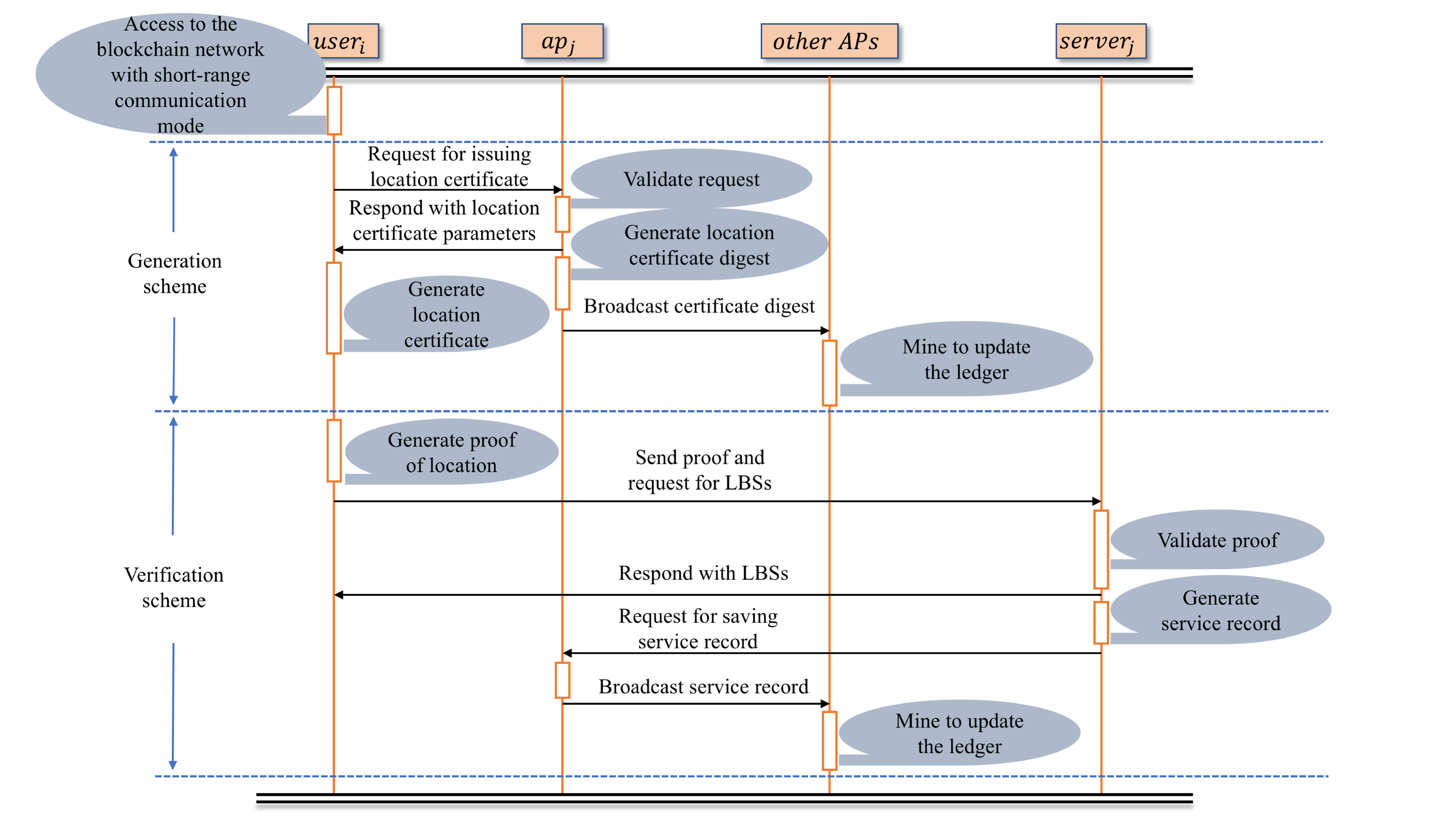}
\caption{Process Diagram of zk-PoL}
\label{process}
\end{figure*}

Focusing on the interaction details among all system components, we propose the zk-PoL protocol. Our protocol consists of two parts: the generation and verification scheme, where generation scheme mainly describes the issuance of the location certificate, and verification scheme mainly describes the generation and verification of the proof. The general process of the protocol is shown in Fig.~\ref{process} and the details are given as follows.
\subsection{Generation scheme}
  Assuming that ${ user }_{ i }$ needs to request for LBSs from the ${server}_{j}$. The first thing he should do is to access to the blockchain network and communicate with the nearby access point $ap_{ j }$ with short-range communication mode, such as WiFi, bluetooth and ZigBee. Then the ${ user }_{ i }$ should request for location certificate issuing from the $ap_{ j }$, and the request is denoted as
\begin{equation}
   { req }_{ i\rightarrow j }:{ \left\{ \begin{matrix} { inf }_{ i }:\left\{ \begin{matrix} { pk }_{ user }^{ i } \\ { \left< longitude,latitude \right>  }_{ user }^{ i } \end{matrix} \right\} , \\ { \left\{ Hash\left( { inf }_{ i } \right)  \right\}  }_{ { sk }_{ user }^{ i } } \end{matrix} \right\}  }_{ { pk }_{ ap }^{ j } },
\end{equation}
where ${ req }_{ i\rightarrow j }$ denotes the request sent by peer $i$ to peer $j$. The ${ sk }_{ user }^{ i }$ denotes the private key of the ${ user }_{ i }$; the ${ pk }_{ user }^{ i }$ and ${ pk }_{ ap }^{ j }$ denote the public key (also the identity in the blockchain network) of ${ user }_{ i }$ and $ap_{ j }$, respectively. It is worth noting that the user equipment has no correspondence with ID in the blockchain network. The blockchain network ID only matches the private key. The ${ \left< longitude,latitude \right>  }_{ user }^{ i }$ denotes the geographic coordinates of ${ user }_{ i }$. The ${ user }_{ i }$ uses ${ sk }_{ user }^{ i }$ to sign the ${inf}_{i}$ to prevent the malicious adversary from tampering with it, and uses ${ pk }_{ ap }^{ j }$ to encrypt the ${req}_{i\rightarrow j}$ to ensure that only the $ap_{ j }$ can correctly decrypt it. 

After receiving ${req}_{i\rightarrow j}$, the $ap_{ j }$ processes it according to the following rules:
\begin{itemize}
    \item If the ${req}_{i\rightarrow j}$ has been tampered with, it will bediscarded.
    \item If the geographic coordinates in ${inf}_{i}$ is out of the range of short-range communication, the ${req}_{i\rightarrow j}$ is discarded.
    \item Otherwise, the $ap_{ j }$ responds to the ${ user }_{ i }$.
\end{itemize}  
The response is defined as
 \vspace*{-0.5\baselineskip}
\begin{equation}
{ { res }_{ j\rightarrow i }:\left\{ \begin{matrix} { inf }_{ j }:\left\{ \begin{matrix} { rand }_{ user }^{ i } \\ { time }_{ user }^{ i } \end{matrix} \right\} , \\ { \left\{ Hash\left( { inf }_{ j } \right)  \right\}  }_{ { sk }_{ ap }^{ j } } \end{matrix} \right\}  }_{ { pk }_{ user }^{ i } },
 \vspace*{-0.5\baselineskip}
\end{equation}
where ${ rand }_{ user }^{ i }$ is a random number and can be regarded as a serial number of certificate. The ${ time }_{ user }^{ i }$ indicates the moment when the certificate is generated. Similarly, the $ap_{ j }$ uses its own private key to sign the ${ inf }_{ j }$, and uses ${ user }_{ i }$'s public key to encrypt the ${ res }_{ j\rightarrow i }$. Meanwhile, with ${ inf }_{ i }$ and ${ inf }_{ j }$, the ${ap}_{ j }$ generates a digest of location certificate ${ dig }_{ i }$
 \vspace*{-0.5\baselineskip}
\begin{equation}\label{dig}
    { dig }_{ i }:\left\{ Hash\left( \begin{matrix} \begin{matrix} { pk }_{ user }^{ i } \\ { \left< longitude,latitude \right>  }_{ user }^{ i } \end{matrix} \\ { rand }_{ user }^{ i },{ time }_{ user }^{ i } \end{matrix} \right)  \right\}.
 \vspace*{-0.5\baselineskip}
\end{equation}
Then, the ${ dig }_{ i }$ is broadcast to miners (APs) in the blockchain network.

The ${ user }_{ i }$ checks whether the ${ res }_{ j\rightarrow i }$ has been tampered with or not. If it is, the ${ user }_{ i }$ discards the ${ res }_{ j\rightarrow i }$. If not, with ${ inf }_{ i }$ and ${ inf }_{ j }$, the ${ user }_{ i }$ generates ${ cer }_{ i }$
 \vspace*{-0.5\baselineskip}
\begin{equation}\label{cer}
    { cer }_{ i }:\left\{ \begin{matrix} \begin{matrix} { pk }_{ user }^{ i } \\ { \left< longitude,latitude \right>  }_{ user }^{ i } \end{matrix} \\ { rand }_{ user }^{ i },{ time }_{ user }^{ i } \end{matrix} \right\}.
 \vspace*{-0.5\baselineskip}
\end{equation}
After ${ cer }_{ i }$ is generated, the ${ap}_{ j }$ will delete ${ inf }_{ i }$ and ${ inf }_{ j }$ according to the embedded smart contract. Meanwhile, the miners package the received ${dig}_{ i }$ into their own blocks. Then by mining, one AP is selected to add its block to the blockchain, updating the ledger. Once ${dig}_{ i }$ is recorded in the ledger, it cannot be tampered with. The ${ user }_{ i }$ can generate a proof proving that his ${ cer }_{ i }$ is in accordance with the ${dig}_{ i }$ in the ledger to obtain related LBSs from the ${ server }_{ j }$.

\subsection{Verification scheme}
Due to the different service requirements of different LBSs, when the ${ user }_{ i }$ requests for LBSs from the ${ server }_{ j }$, he needs to choose different input parameters, as shown in TABLE \uppercase\expandafter{\romannumeral1}, to generate the appropriate proof proving that he satisfies the service requirement. 
\begin{table}[ht]
\label{input parameters}
\caption{input parameters}
\begin{center}
\begin{tabular}{c|c|c}
\hline
\textbf{Level}& \textbf{Public Parameters} & \textbf{Private Parameters}  \\
\hline
\textbf{1} &$null$ & \tabincell{c}{${ pk }_{ user }^{ i }$, \\${ \left< longitude,latitude \right>  }_{ user }^{ i }$, \\${ rand }_{ user }^{ i }$, \\${ time }_{ user }^{ i }$}\\
\hline
\textbf{2} &${ \left< longitude,latitude \right>  }_{ user }^{ i }$ & \tabincell{c}{${ pk }_{ user }^{ i }$, \\${ rand }_{ user }^{ i }$, \\${ time }_{ user }^{ i }$}\\
\hline
\textbf{3} &\tabincell{c}{${ \left< longitude,latitude \right>  }_{ user }^{ i }$,\\${ time }_{ user }^{ i }$} & \tabincell{c}{${ pk }_{ user }^{ i }$, \\${ rand }_{ user }^{ i }$}\\
\hline
\textbf{4} &\tabincell{c}{${ pk }_{ user }^{ i }$,\\${ \left< longitude,latitude \right>  }_{ user }^{ i }$,\\${ time }_{ user }^{ i }$} & ${ rand }_{ user }^{ i }$\\
\hline
\end{tabular}
\label{tab1}
\end{center}
\end{table}

The four levels of input parameters expose different necessary information to the server, achieving so-called hierarchical privacy protection, which can be applied to different scenarios:
\begin{itemize}
    \item Level 1: The ${ user }_{ i }$ locally generates a proof with input parameters in level 1. This can be used to prove that someone has appeared at certain time in the range covered by APs and be applied to the scenarios such as authorization of attractions review where the critic must prove that he has been some tourist area in order to get commenting permission. 
    \item Level 2: The ${ user }_{ i }$ locally generates a proof with input parameters in level 2, disclosing its own coordinate. This can be used to prove that someone has appeared at a particular location and certain moment, and can be applied to the scenarios such as fixed-point Punching for Coupon Service (PCS) where the user needs to punch in the specific place in order to get the coupon.
    \item Level 3: The ${ user }_{ i }$ locally generates a proof with input parameters in level 3, disclosing its own coordinate and time. This can be used to prove that someone has appeared at a particular location and particular moment, and can be applied to scenarios such as real-time location based recommendations where the user must prove that he is now at a specific location in order to get precise recommendations.
    \item Level 4: The ${ user }_{ i }$ locally generates a proof with input parameters in level 4, disclosing its own coordinate, time, and identity, i.e. ${ pk }_{ user }^{ i }$. This can be used to prove that a particular person has appeared at a particular place and a particular moment. For example, a suspect needs the alibi to prove that he is not at the crime scene at a particular moment.
\end{itemize}

Without loss of generality, the following takes the PCS as an example to illustrate the details of the verification scheme. In PCS, the ${ user }_{ i }$ needs to punch in the specific place to get the coupon. Note that the ID of ${ user }_{ i }$ when he requests LBSs from the ${ server }_{ j }$ is no longer ${ pk }_{ user }^{ i }$, so the ${ user }_{ i }$'s identity is not exposed. The ${ user }_{ i }$ should use the location certificate ${ cer }_{ i }$ to generate a relevant proof and then send it to the ${ server }_{ j }$ for PCS. The specific process goes as following.
\subsubsection{Computation to QAP}
As mentioned earlier, any complex computation can be converted into QAP. The complex computations in the PCS case is
\begin{equation}
\left\{
\begin{matrix} Hash({ rand }_{ user }^{ i })={ hr }_{ i } \\ Hash\left( pubpara,pripara \right) ={ dig }_{ i } \end{matrix}
\right.,
\end{equation}
where ${ hr }$ is the hash of serial number and it is set for special single-use service. Its specific role will be discussed below. The $pubpara$ and $pripara$ are public parameters and private parameters in TABLE \uppercase\expandafter{\romannumeral1}, respectively. The QAP obtained from computation is
 \vspace*{-0.5\baselineskip}
\begin{equation}
    \left\{ \left\{ { l }_{ i }\left( x \right) ,{ r }_{ i }\left( x \right) ,{ o }_{ i }\left( x \right)  \right\} ,t\left( x \right)  \right\} \quad i\in \left\{ 0,...m,...,n \right\} ,
 \vspace*{-0.5\baselineskip}
\end{equation}
where ${ l }_{ i }\left( x \right) $, ${ r }_{ i }\left( x \right) $ and ${ o }_{ i }\left( x \right)$ are variable polynomials; $n$ is the number of variables; ${ l }_{ 0 }\left( x \right) $, ${ r }_{ 0 }\left( x \right) $ and ${ o }_{ 0 }\left( x \right) $ are polynomials of pseudo-variable $one$ and ${ l }_{ 1 }\left( x \right) $ to ${ l }_{ m }\left( x \right) $, ${ r }_{ 1 }\left( x \right) $ to ${ r }_{ m }\left( x \right) $ and ${ o }_{ 1 }\left( x \right) $ to ${ o }_{ m }\left( x \right) $ are polynomials of public variables. In this case $m=1$, which represents the public parameter ${ \left< longitude,latitude \right>  }_{ user }^{ i }$; $t\left( x \right) $ is the target polynomial and its degree $d$ denotes the number of operations. 
\subsubsection{Setup}
After the conversion to QAP is completed, some randoms $\left< s,g,{ \rho  }_{ l },{ \rho  }_{ r },\alpha _{ l },\alpha _{ r },{ \alpha  }_{ o },\beta ,\gamma  \right>$ are needed. With these randoms, the APs generate proving key
 \vspace*{-0.5\baselineskip}
\begin{equation}
    proving\quad key:\left\{ \begin{matrix} \left\{ { g }^{ { s }^{ i } } \right\}  \\ \begin{matrix} \left\{ { g }_{ l }^{ { l }_{ j }\left( s \right)  },{ g }_{ r }^{ { r }_{ j }\left( s \right)  },{ g }_{ o }^{ { o }_{ j }\left( s \right)  } \right\}  \\ \begin{matrix} \left\{ { g }_{ l }^{ { { \alpha  }_{ l }l }_{ j }\left( s \right)  },{ g }_{ r }^{ { { \alpha  }_{ r }r }_{ j}\left( s \right)  },{ g }_{ o }^{ { { \alpha  }_{ o }o }_{ j }\left( s \right)  } \right\}  \\ \left\{ { g }_{ l }^{ { { \beta  }l }_{j }\left( s \right)  }{ g }_{ r }^{ { { \beta  }r }_{ j }\left( s \right)  }{ g }_{ o }^{ { { \beta  }o }_{ j }\left( s \right)  } \right\}  \end{matrix} \end{matrix} \end{matrix} \right\},
 \vspace*{-0.5\baselineskip}
\end{equation}
where ${ g }_{ l }={ g }^{ { \rho  }_{ l } }$, ${ g }_{ r }={ g }^{ { \rho  }_{ r } }$, ${ g }_{ o }={ g }^{ { \rho  }_{ l }\cdot { \rho  }_{ r } }$, $i\in \left\{ 1,...,d \right\} $ and $j\in \left\{ m+1,...,n \right\} $, and verification key
 \vspace*{-0.5\baselineskip}
\begin{equation}
    verification\quad key:\left\{ \begin{matrix} { g }^{ 1 },{ g }_{ o }^{ t\left( s \right)  } \\ \left\{ { { g } }_{ l }^{ { l }_{ i }\left( s \right)  },{ { g } }_{ r }^{ { r }_{ i }\left( s \right)  },{ { g } }_{ o }^{ { o }_{ i }\left( s \right)  } \right\}  \\ { g }^{ { \alpha  }_{ l } },{ g }^{ { \alpha  }_{ r } },{ g }^{ { \alpha  }_{ o } },{ g }^{ \gamma  },{ g }^{ \gamma \beta  } \end{matrix} \right\},
 \vspace*{-0.5\baselineskip}
\end{equation}
where $i\in \left\{ 0,...,m \right\} $.

Once proving key and verification key are generated, these randoms are deleted by APs. Any adversary with these randoms has the ability to falsify a false proof to cheat the server. This is why these randoms are called 'toxic waste'.

The above two phases can be collectively referred to as Trusted Setup (TS) for generating CRS, i.e., the pair of proving and verification keys. In our protocol, TS is executed by trusted APs at the system initialization.
\subsubsection{Calculate witness}
In order to generate the proof, the ${ user }_{ i }$ needs to calculate all the intermediate variables $witness$
 \vspace*{-0.5\baselineskip}
\begin{equation}
    witness:\left\{ { v }_{ m+1 },{ ...,v }_{ n } \right\}.
 \vspace*{-0.5\baselineskip}
\end{equation}
\subsubsection{Generate proof}
The ${ user }_{ i }$ calculates polynomials $L\left( x \right)$, $R\left( x \right)$ and $O\left( x \right) $ through assigning ${ v }_{ i }$ to ${ l }_{ i }\left( x \right)$, ${ r }_{ i }\left( x \right)$ and $o_{ i }\left( x \right)$, as shown in \eqref{assign}, and then obtains $h\left( x \right) $
 \vspace*{-0.5\baselineskip}
\begin{equation}
    h\left( x \right) =\frac { L\left( x \right) R\left( x \right) -O\left( x \right)  }{ t\left( x \right)  }.
 \vspace*{-0.5\baselineskip}
\end{equation}
Then using the proving key, the ${ user }_{ i }$ calculates
\begin{equation}
    { g }_{ l }^{ { L }_{ p }\left( s \right)  }={ g }_{ l }^{ \sum _{ i=m+1 }^{ n }{ { v }_{ i }{ l }_{ i }\left( s \right)  }  },
 \vspace*{-0.5\baselineskip}
\end{equation}
\begin{equation}
    { g }_{ l }^{ { L }_{ p }^{ ' }\left( s \right)  }={ g }_{ l }^{ { \alpha  }_{ l }\sum _{ i=m+1 }^{ n }{ { v }_{ i }{ l }_{ i }\left( s \right)  }  }.
\end{equation}
Similarly, ${ g }_{ r }^{ R_{ p }\left( s \right)  }$, ${ g }_{ r }^{ R_{ p }^{ ' }\left( s \right)  }$, ${ g }_{ o }^{ O_{ p }\left( s \right)  }$ and ${ g }_{ o }^{ O_{ p }^{ ' }\left( s \right)  }$ can be obtained, and the variable consistency polynomial is calculated by
 \vspace*{-0.5\baselineskip}
\begin{equation}
    \begin{aligned}
        { g }^{ Z\left( s \right)  }=&\prod _{ i=m+1 }^{ n }{ { { g }_{ l } }^{ { v }_{ i }\beta { l }_{ i }\left( s \right)  } } { { g }_{ r } }^{ { v }_{ i }\beta r_{ i }\left( s \right)  }{ { g }_{ o } }^{ { v }_{ i }\beta { o }_{ i }\left( s \right)  }\\ =&{ { g }_{ l } }^{ \beta { L }_{ p }\left( s \right)  }{ { g }_{ r } }^{ \beta { R }_{ p }\left( s \right)  }{ { g }_{ o } }^{ \beta { O }_{ p }\left( s \right)  }
    \end{aligned}
 \vspace*{-0.5\baselineskip}
\end{equation}
In this way, the ${ user }_{ i }$ obtains the proof ${ pro }_{ i }$
 \vspace*{-0.5\baselineskip}
\begin{equation}
    { pro }_{ i }:\left\{ \begin{matrix} { { g }_{ l }^{ L_{ p }\left( s \right)  },{ g }_{ l }^{ { L }_{ p }^{ ' }\left( s \right)  }, }{ g }_{ o }^{ R_{ p }\left( s \right)  },{ g }_{ r }^{ R_{ p }^{ ' }\left( s \right)  }, \\ { { g }_{ o }^{ O_{ p }\left( s \right)  },{ g }_{ o }^{ { O }_{ p }^{ ' }\left( s \right)  } },{ g }^{ Z\left( s \right)  },{ g }^{ h\left( s \right)  } \end{matrix} \right\}.
 \vspace*{-0.5\baselineskip}
\end{equation}
After that, the ${ user }_{ i }$ sends proof to the ${ server }_{ j }$ and requests the PCS from the ${ server }_{ j }$, and the request is
 \vspace*{-0.5\baselineskip}
\begin{equation}
    { req }_{ i\rightarrow j }:\left\{ { pro }_{ i },{ ind }_{ PCS },{ hr }_{ i } \right\},
 \vspace*{-0.5\baselineskip}
\end{equation}
where ${ ind }_{ PCS }$ is the index of PCS which ${ user }_{ i }$ requested.
\subsubsection{Verify proof}
The ${ server }_{ j }$ checks if the service record has already existed in the ledger. If it is, the request for PCS is denied. If not, using the verification key, the ${ server }_{ j }$ validate ${ pro }_{ i }$ by checking
 \vspace*{-0.5\baselineskip}
\begin{equation}\label{varible polynomials restriction check}
    \left\{
    \begin{matrix} \begin{matrix} e\left( { g }_{ l }^{ L_{ p }\left( s \right)  },{ g }^{ { \alpha  }_{ l } } \right) =e\left( { g }_{ l }^{ { L }_{ p }^{ ' }\left( s \right)  },g \right)  \\ e\left( { g }_{ r }^{ R_{ p }\left( s \right)  },{ g }^{ { \alpha  }_{ r } } \right) =e\left( { g }_{ r }^{ { R }_{ p }^{ ' }\left( s \right)  },g \right)  \end{matrix} \\ e\left( { g }_{ o }^{ O_{ p }\left( s \right)  },{ g }^{ { \alpha  }_{ o } } \right) =e\left( { g }_{ o }^{ { O }_{ p }^{ ' }\left( s \right)  },g \right)  \end{matrix}
    \right.,
\end{equation}
\begin{equation}\label{variable values consistency check}
    e\left( { g }_{ l }^{ L_{ p }\left( s \right)  }{ g }_{ r }^{ R_{ p }\left( s \right)  }{ g }_{ o }^{ O_{ p }\left( s \right)  },{ g }^{ \beta \gamma  } \right) =e\left( { g }^{ Z\left( s \right)  },{ g }^{ \gamma  } \right), 
\end{equation}
 \vspace*{-0.5\baselineskip}
\begin{equation}\label{valid operations check}
   \begin{matrix} e\left( { g }_{ l }^{ L_{ v }\left( s \right) +L_{ p }\left( s \right)  },{ g }_{ r }^{ R_{ v }\left( s \right) +R_{ p }\left( s \right)  } \right)  \\ =e\left( { g }_{ o }^{ t\left( s \right)  },{ g }^{ h\left( s \right)  } \right) \cdot e\left( { g }_{ o }^{ O_{ v }\left( s \right) +O_{ p }\left( s \right)  },g \right)  \end{matrix},
 \vspace*{-0.5\baselineskip}
\end{equation}
where ${ g }_{ l }^{ L_{ v }\left( s \right)  }={ g }_{ l }^{ { l }_{ 0 }\left( s \right) +...+{ v }_{ m }{ l }_{ m }\left( s \right)  }$, and similarly for ${ g }_{ r }^{ R_{ v }\left( s \right)  }$ and ${ g }_{ o }^{ O_{ v }\left( s \right)  }$. After that, the ${ server }_{ j }$ provides PCS for the ${ user }_{ i }$ and sends a service record ${ rec }_{ ji }$ to the ${ ap }_{ j }$, which is defined as 
 \vspace*{-0.5\baselineskip}
\begin{equation}
    \begin{matrix} { rec }_{ ji }:\left\{ { pk }_{ sever }^{ j },{ ind }_{ PCS },{ hr }_{ i } \right\}.
    \end{matrix}
 \vspace*{-0.5\baselineskip}
\end{equation}
Then $ { rec }_{ ji }$ is broadcasted to other APs in the blockchain network and saved in the ledger through mining, which indicates that the ${ server }_{ j }$ has provided PCS for the ${ user }_{ i }$. When the ${ user }_{ i }$ attempts to request PCS again to obtain another coupon, the ${ server }_{ j }$ can reject the request because the service record has already existed in the ledger.

At this point, we have introduced the complete zk-PoL with input parameters in level 1. The zk-PoL with input parameters in other levels has some subtle differences in generating CRS, calculating witness, generating proof and verifying proof, but the general process is the same. Because of space limitation, the details will not be discussed here.

\section{Evaluation}
In this section, we analyze the proposed system security to resist main attacks, including cheating on certificate parameters, cheating on service and privacy inference attack. Moreover, we will also discuss computational efficiency according to the experiment.
\subsection{Security}
\subsubsection{Cheating on Certificate Parameters}
In generation scheme, the certificate parameters ${ time }$ and ${ rand }$ are generated by the trusted APs, so these two parameters can not be falsified. A malicious user can only cheat on ${ \left< longitude,latitude \right>  }_{ user }$, and ${ pk_{ user } }$. However, by detecting whether the coordinate sent by the malicious peer is in the range of short-range communication, the location cheating can be prevented. In addition, to prevent cheating on ${ pk_{ user } }$, the AP checks
 \vspace*{-0.5\baselineskip}
\begin{equation}
    { D }_{ { pk }_{ user }^{ ' } }\left( { E }_{ { sk }_{ user } }\left( Hash\left( inf \right)  \right)  \right) =Hash\left( inf \right),
 \vspace*{-0.5\baselineskip}
\end{equation}
where $D$ and $E$ are encryption algorithms and decryption algorithms of asymmetric cryptography, respectively. The ${ pk }_{ user }^{ ' } $ is the ID to be checked.

\subsubsection{Cheating on Services}
In the verification scheme, the server, by checking variable polynomials restriction, variable values consistency and valid operations, checks the validity of $pro$, as shown in \eqref{varible polynomials restriction check}, \eqref{variable values consistency check} and \eqref{valid operations check}. In addition, by deleting 'toxic waste', it can be ensured that any malicious user without certificate parameters can not forge $pro$ for corresponding services. Hence cheating on services can be avoided. Moreover, to prevent cheating on obtaining special one-off services repeatedly, the server checks
 \vspace*{-0.5\baselineskip}
\begin{equation}
    \left( { pk }_{ server },ind,hr\right) \in \left\{ rec \right\},
 \vspace*{-0.5\baselineskip}
\end{equation}
where $\left\{ rec \right\} $ denotes the set of service records stored in the ledger.

\subsubsection{Privacy Inference Attack}
In the generation scheme, the private information of the user is only available to the user himself and the APs. The APs are prevented by the embedded smart contract from disclosing the user's privacy. Also, the digest of the location certificate stored in the public ledger does not reveal the user's privacy to any malicious peer.

In the verification scheme, according to the characteristics of the trapdoor function, it is hard to get the $witness:\left\{ { v }_{ m+1 }{ ,...,v }_{ n } \right\}$ from CRS and $pro$, so our zk-PoL only provides the necessary private information for the server without revealing any other privacy.

\subsection{Computational Efficiency}
\begin{figure}[ht]
\centering
\includegraphics[width=8.5cm]{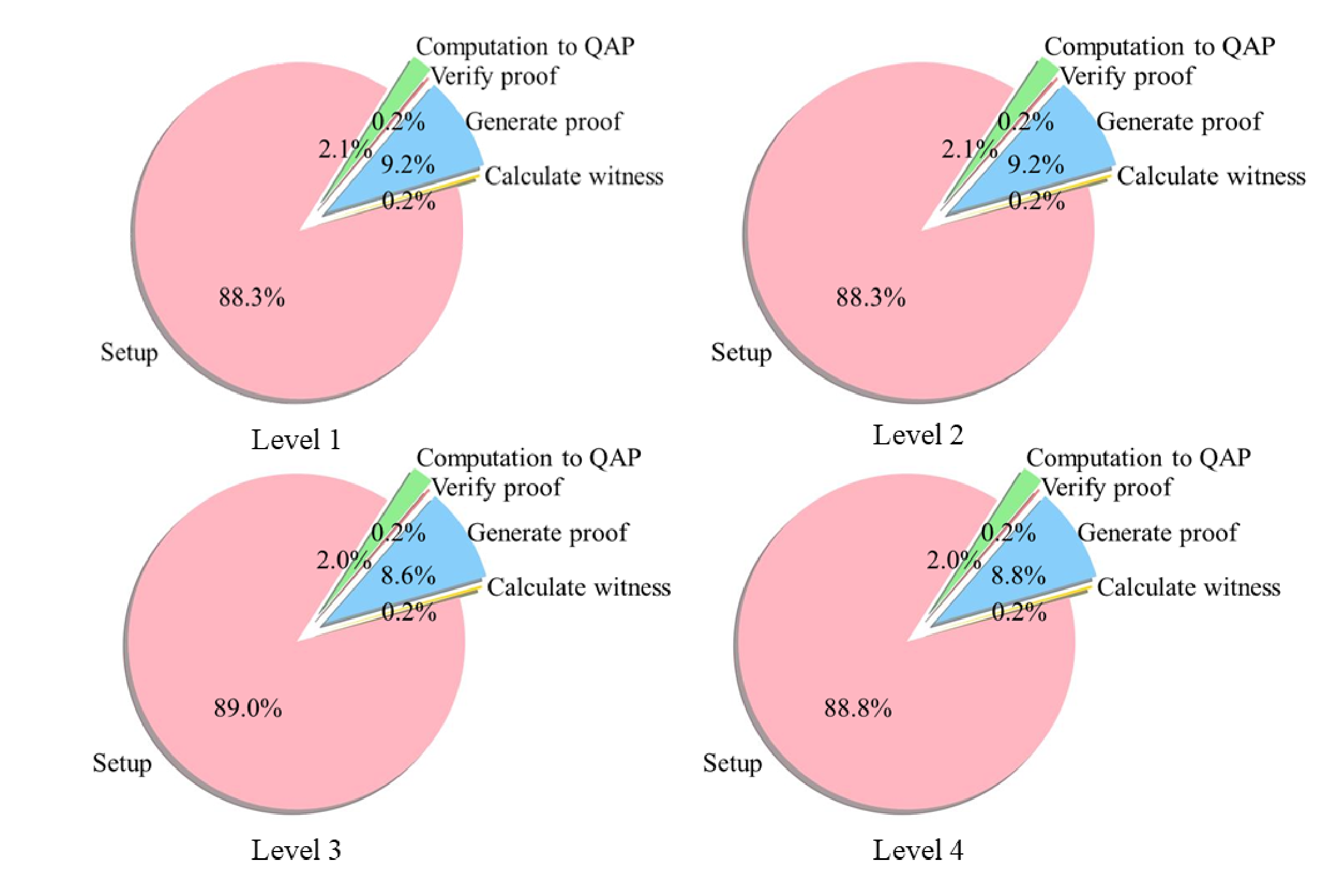}
\caption{Distribution of Average Execution Time}
\label{timeofeachpart}
\vspace*{-0.5\baselineskip}
\end{figure}

Our experiment is on MateBook 13 with 1.8Ghz i7 CPU and 8G RAM. Moreover, the experiment is based on Circom and Snarkjs\cite{QAPcompiler,snarkjs}. Circom is a circuit generator to translate program executions to arithmetic circuits. Snarkjs is a JavaScript based  cryptographic proof system for verifying the satisfiability of arithmetic circuits generated by Circom. We experiment with parameters in four levels, and the percentage of average execution time for each phase is as shown in Fig.~\ref{timeofeachpart}. 
\vspace*{-0.5\baselineskip}
\begin{itemize}
    \item \emph{Computation to QAP:} The percentages of this phase in four levels are 2.1\%, 2.1\%, 2.0\%, and 2.0\%, respectively. 
    \item \emph{Setup:} The percentages of this phase in four levels are 88.3\%, 88.3\%, 89.0\%, and 88.8\%, respectively.
    \item \emph{Calculate witness:} The percentages of this phase in four levels all are 0.2\%.
    \item \emph{Generate proof:} The percentages of this phase in four levels are 9.2\%, 9.2\%, 8.6\%, and 8.8\%, respectively.
    \item \emph{Verify proof:} The percentages of this phase in four levels all are 0.2\%.
\end{itemize}

The result shows that the distributions of average execution time in different levels of zk-PoL are basically the same. Moreover, due to the complexity of the elliptic curve point addition and hash algorithm, the percentage of the average execution time of TS is large, accounting for about 90\% of the total execution time. Fortunately, the TS only needs to be executed once during the system initialization. Therefore, we only need to consider the execution time of the three phases: \emph{Calculate witness}, \emph{Generate proof} and \emph{Verify proof}.

\begin{figure}[ht]
\centering
\includegraphics[width=8.5cm]{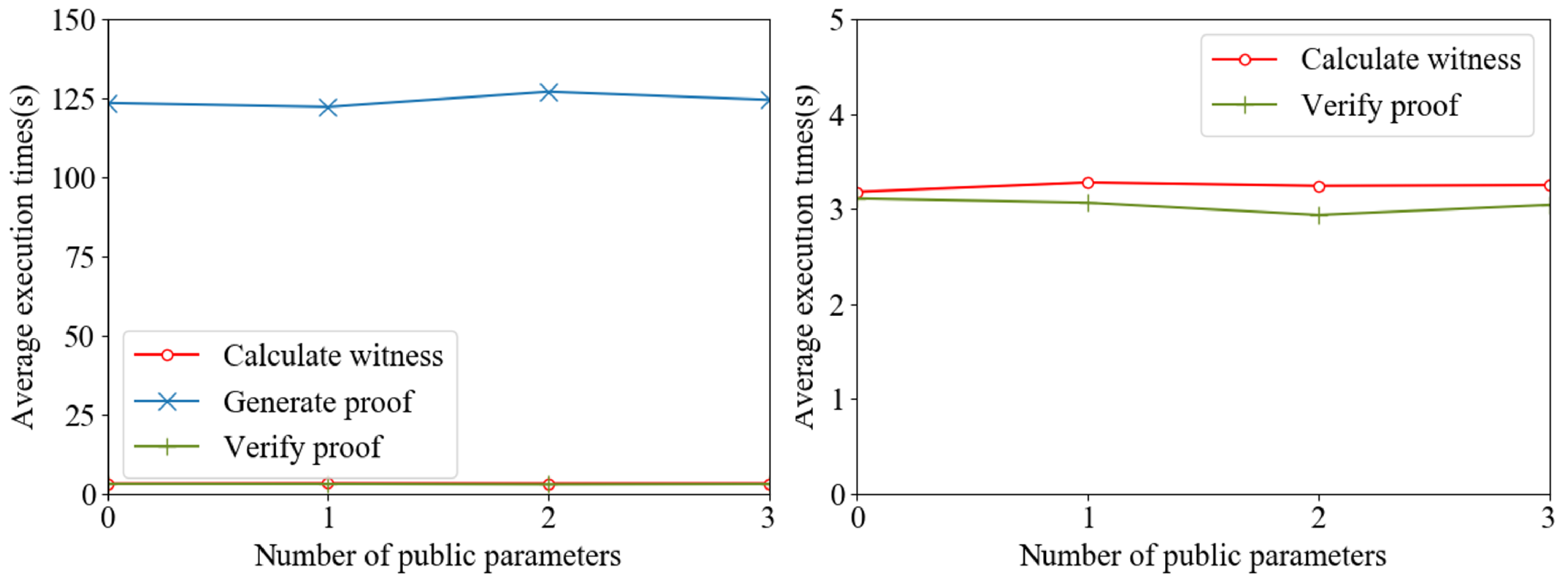}
\caption{Average Execution Time of Four Levels}
\label{averageexecutiontime}
\end{figure}

 The relationships between the average execution time and the number of public parameters in the three phases are shown in the Fig.~\ref{averageexecutiontime}. It can be seen that the average execution time of each phase across different public parameters is basically the same. In addition, the total execution time of the three phases that need to be repeated during different services is about 2 minutes. This execution time is acceptable in some delay-tolerant LBSs, such as the PCS case.
\section{Conclusion}
In this paper, we bring zero-knowledge proof to LBSs for the first time, proposing zk-PoL protocol which consists of generation scheme and verification scheme. This protocol allows a user to achieve hierarchical, user-controllable privacy protection by choosing different input parameters. In addition, for main attacks, we analyze the security of our system, and the result shows that the system has excellent security against these attacks. Finally, we experiment with our system, demonstrating that the performance is independent of the privacy protection level. In diverse services, the average execution time of the phases needed to be repeated (all phases except the TS phase) is about 2 minutes, and the ratio of that to the total time cost is about 10\%. This performance is appropriate to delay-tolerant LBSs, but not to delay-sensitive LBSs. In our future work, we will implement different circuit generators and cryptographic proof systems \cite{184425} to improve the system performance, making our zk-PoL appropriate to more kinds of LBSs.



%
\bibliographystyle{IEEEtran}
\bibliography{ref}

\end{document}